\documentclass[envcountsame]{llncs}
\usepackage{amsmath}
\usepackage{latexsym}
\usepackage{graphicx}
\usepackage{pstricks}
\usepackage{amsfonts}
\usepackage{amssymb}

\title{A Correspondence between Phenomenological and $IBM-1$ Models of Even Isotopes of $Yb$}
\author{Abdurahim Okhunov\inst{1,4} \and Fadhil I. Sharrad\inst{2} \and Anwer A. Al-Sammarea\inst{3,5} }
\institute{Department of Science in Engineering, Faculty of Engineering\\
               International Islamic University Malaysia \\
              P.O. Box 10, 50728 Kuala Lumpur, Selangor, Malaysia \\
              \email{abdurahimokhun@iium.edu.my}
\and
           Department of Physics, College of Science
           University of Kerbala, Karbala, Iraq \\
\and
           Department of Chemistry, College of Education, Faculty of Science  \\
           University of Samara, Selah, Iraq \\
\and
           Institute of Nuclear Physics, Academy of Science Republic of Uzbekistan \\
           100214 Ulugbek, Tashkent, Uzbekistan \\
\and
           Quantum Science Centre, Department of Physics, Faculty of Science \\
           University of Malaya, 50603 Kuala Lumpur, Malaysia \\
}

\date{Received: date / Accepted: date}

\usepackage{setspace}
\begin{document}
\maketitle
\begin{abstract}
This paper studies the nonterminal complexity of weakly
conditional grammars, and it proves that every recursively
enumerable language can be generated by a weak conditional grammar
with no more than \textit{seven} nonterminals. Moreover, it is
shown that the number of nonterminals in weakly conditional
grammars without erasing rules leads to an infinite hierarchy of
families of languages generated by weakly conditional grammars.

Energy levels and the reduced probability of $E2$-- transitions
for Ytterbium isotopes with proton number $Z=70$ and neutron
numbers between 100 and 106 have been calculated through
phenomenological ($PhM$) and interacting boson ($IBM-1$) models.
The predicted low-lying levels (energies, spins and parities) and
the reduced probability for $E2$-- transitions results are reasonably
consistent with the available experimental data. The predicted
low-lying levels ($gr$--, $\beta_1$-- and $\gamma_1$-- band) by
produced in the $PhM$ are in good agreement with the experimental
data comparison with those by $IBM-1$ for all nuclei of interest.
In addition, the phenomenological model was successful in
predicted the $\beta_2$--, $\beta_3$--, $\beta_4$--, $\gamma_2$--
and $1^+$-- band while it was a failure with $IBM-1$. Also, the
$3^+$-- band is predicted by the $IBM-1$ model for $^{172}Yb$ and
$^{174}Yb$ nuclei. All calculations are compared with the
available experimental data.
\end{abstract}

{{\bf Keywords \ \ :} \ Ytterbium ($Yb$); Energy levels; model;
even-even; isotopes; nuclei}


%%\keywords{Energy levels, model, even - even, isotopes, nuclei}

%%\pacs{21.10.-k, 21.10.Ky, 21.10.Hw}

%\begin{document}
%\maketitle

\section{Introduction}

The medium-to heavy-mass Ytterbium ($Yb$) isotopes located in the
rare-earth mass region are well-deformed nuclei that can be
populated to very high spin. Much experimental information on
even-odd-mass of $Yb$ isotopes has become more abundant
\cite{Mo87}-\cite{Granja05}. For the heavier $A=174$ to $178$
nuclei \cite{Lee97}, previous work using deep inelastic reactions
and Gammasphere have begun to reveal much information about the
high-spin behavior of these neutron-rich $Yb$ isotopes. The yrast
states in the well deformed rare-earth region have been described
by using the projected shell model \cite{Hara80}-\cite{Archer98}.

Prior to the present work the level structure of ground band state
and low-lying excited states of even-even nuclei has been studied
both theoretically and experimentally
\cite{Gelletly87}, including the  decay, Coulomb
excitation and various transfer reactions.

In this study, two calculations for energy levels of
$^{170,172,174,176}Yb$ isotopes have been done by using two
different models phenomenological model ($PhM$), and interacting
boson model ($IBM-1$). Positive parity state energies and the
reduced probability of E2 transitions are calculated and compared
with the available experimental data. The structure of excited
levels is discussed.

\section{Theoretical models}

The calculations have been performed by using the phenomenological
and interacting boson models. In the next subsection, we will
explain these models.

\subsection{Phenomenological model ($PhM$)}

To analyze the properties of low-lying positive parity states in
$Yb$ isotopes, the $PhM$ of \cite{Usmanov97,Usmanov10} is used.
This model takes into account the mixing of states of the $gr$--,
$\beta$--, $\gamma$-- and $K^{\pi}=1^+$-- band. The model
Hamiltonian is chosen in the following form
\begin{eqnarray}\label{Eq.1}
H=H_{rot}(I^2)+ H_{K,K'}^{\sigma}
\end{eqnarray}

\noindent here $H_{rot}(I^2$-- is rotational part of Hamiltonian,

\begin{eqnarray}\label{Eq.2}
H_{_{K'K}}^{\sigma}(I)=\omega_{_K}\delta_{_{K,K'}}-\omega_{rot}(I)(j_x)_{_{K,K'}}\xi(I,K)\delta_{_{K,K'\pm1}}
\end{eqnarray}

\noindent where $\omega_{_K}$-- bandhead energy of rotational
band, $\omega_{rot}(I)$-- is the rotational frequency of core,
$(j_x)_{_{K,K'}}$-- is the matrix elements which describe Coriolis
mixture between rotational bands:
\begin{eqnarray}
(j_x)_{_{gr,1}}=-\sqrt{3}\cdot\tau_0,
(j_x)_{_{\beta,1}}=-\sqrt{3}\cdot\tau_1,
(j_x)_{_{\gamma,1}}=-1\cdot\tau_2 \nonumber
\end{eqnarray}
 and
\begin{eqnarray}
\xi(I,0)=1 \ \ \ \ \ \xi(I,2)=\sqrt{1-\frac{2}{I(I+1)}} \nonumber
\end{eqnarray}

The eigenfunction of Hamiltonian model (\ref{Eq.1}) has the form
\begin{eqnarray}\label{Eq.3}
|IMK>&=&\left[\frac{2I+1}{16\pi^2}\right]^{\frac{1}{2}}\left\{\sqrt{2}\Psi_{gr,K}^ID_{MK}^{I}(\theta)\right.
\\ \nonumber
&+&\left.\sum_{K'}\frac{\Psi_{K',K}^I}{\sqrt{1+\delta_{K',0}}}
\left[D_{M,K'}^I(\theta)b_{K'}^++(-1)^{I+K'}D_{M,-K'}^I(\theta)b_{-K'}^+\right]\right\}|0>
\nonumber
\end{eqnarray}

\noindent here $\Psi_{K',K}^{I}$ is the amplitude of mixture of
basis states.

Solving the Schr\"{o}dinger equation one can determine the
eigenfunctions and eigenenergies of the positive parity states.
\begin{eqnarray}\label{Eq.4}
\left(H_{K,\nu}^I-\varepsilon_{\nu}^{I}\right)\Psi_{K,\nu}^{I}=0
\end{eqnarray}

\noindent we can determined the eigenfunctions and eigenenergies
of the positive--parity states.

 The complete energy of a state is given by
\begin{eqnarray}\label{Eq.5}
E_{\nu}^I(I)=E_{rot}(I)+\varepsilon_{\nu}^I(I)
\end{eqnarray}

The rotating-core energy $E_{rot}(I)$ is calculated by using the
Harris parameterizations \cite{Harris65} of the energy and the
angular momentum, that is

\begin{eqnarray}\label{Eq.6}
E_{rot}(I)=\frac{1}{2}\jmath_{_0}\omega_{rot}^2(I)+\frac{3}{4}\jmath_{_1}\omega_{rot}^4(I)
\end{eqnarray}
\begin{eqnarray}\label{Eq.7}
\sqrt{I(I+1)}=\jmath_{_0}\omega_{rot}(I)+\jmath_{_1}\omega_{rot}^3(I)
\end{eqnarray}

\noindent where $\jmath_{_0}$ and $\jmath_{_1}$ are the inertial
parameters of the rotational core.

The rotational frequency of the core $\omega_{rot}(I)$ is found by
solving the cubic equation (\ref{Eq.7}). This equation has two
imaginary roots and one real root. The real root is
\begin{footnotesize}
\begin{eqnarray}\label{Eq.8}
\omega_{rot}(I)=
\left\{\frac{\widetilde{I}}{2\jmath_1}+\left[\left(\frac{\jmath_{_0}}{3\jmath_{_1}}\right)^3+\left(\frac{\widetilde{I}}{2\jmath_{_1}}\right)^2
\right]^{\frac{1}{2}}\right\}^{\frac{1}{3}} \\ \nonumber
+\left\{\frac{\widetilde{I}}{2\jmath_{_1}}-\left[\left(\frac{\jmath_{_0}}{3\jmath_{_1}}\right)^3+\left(\frac{\widetilde{I}}{2\jmath_{_1}}\right)^2
\right]^{\frac{1}{2}}\right\}^{\frac{1}{3}}
\end{eqnarray}
\end{footnotesize}

\noindent where $\widetilde{I}=\sqrt{I(I+1)}$. Equation
(\ref{Eq.8}) gives $\omega_{rot}(I)$ at the given spin $I$ of the
core.

\subsection{Interacting boson model ($IBM-1$)}

The $IBM$ has become one of the most intensively used nuclear
models, due to its ability to describe the low-lying collective
properties of nuclei across an entire major shell with a
Hamiltonian. In the $IBM-1$ the spectroscopies of low-lying
collective properties of even-even nuclei are described in terms
of a system of interacting $s$ bosons ($L=0$) and $d$ bosons
($L=2$). Furthermore, the model assumes that the structure of
low-lying levels is dominated by excitations among the valence
partials outside major closed shells. In the particle space the
number of proton bosons $N_{\pi}$ and neutron bosons
$N_{\vartheta}$ is counted from the nearest closed shell, and the
resulting total boson number is a strictly conserved quantity. The
underlying structure of the six-dimensional unitary group $SU(6)$
of the model leads to a Hamiltonian, capable of describing the
three specific types of collective structures with classical
geometrical analogues (vibrational \cite{Arima76}, rotational
\cite{Arima78}, and $\gamma$-- unstable \cite{Arima79}) and also
the transitional nuclei \cite{Scholyen78} whose structures are
intermediate. The $IBM-1$ Hamiltonian can be expressed as
\cite{Arima79}

\begin{eqnarray}\label{Eq.9}
H&=&\varepsilon_{_s}(s^+\widetilde{s})+\varepsilon_{_d}\left(d^+\widetilde{d}\right)
\nonumber \\
&+&\sum_{L=0,2,4}\frac{1}{2}(2L+1)^{\frac{1}{2}}C_{_L}
\left[\left(d^+\times
d^+\right)^{(L)}\left(\widetilde{d}\times\widetilde{d}\right)^{(L)}\right]^{(0)}
\nonumber  \\
&+&\frac{1}{2}\widetilde{\vartheta}_{0}\left[\left(d^+\times
d^+\right)^{(0)}\left(\widetilde{s}\times\widetilde{s}\right)^{(0)}
+\left(s^+\times
s^+\right)^{(0)}\left(\widetilde{d}\times\widetilde{d}\right)^{(0)}\right]^{(0)}
\\
&+&\frac{1}{\sqrt{2}}\widetilde{\vartheta}_{2}\left[\left(d^+\times
d^+\right)^{(2)}\left(\widetilde{d}\times\widetilde{s}\right)^{(2)}
+\left(d^+\times
s^+\right)^{(2)}\left(\widetilde{d}\times\widetilde{d}\right)^{(2)}\right]^{(0)}
\nonumber \\
&+&u_2\left[\left(d^+\times
s^+\right)^{(2)}\left(\widetilde{d}\times\widetilde{s}\right)^{(2)}\right]^{(0)}
+\frac{1}{2}u_0\left[\left(s^+\times
s^+\right)^{(0)}\left(\widetilde{s}\times\widetilde{s}\right)^{(0)}\right]^{(0)}
\nonumber
\end{eqnarray}

\noindent where $\left(s^{\dag}, d^{\dag}\right)$ and $\left(s,
d\right)$ are creation and annihilation operators for $s$ and $d$
bosons, respectively.

 The $IBM-1$ Hamiltonian equation (\ref{Eq.9}) can be written in general form as \cite{Casten88}
\begin{eqnarray}\label{Eq.10}
H=\varepsilon n_{_d}+a_{0}P^{\dag}
P+a_{1}LL+a_{2}QQ+a_{3}T3T3+a_{4}T4T4
\end{eqnarray}

The full Hamiltonian $H$ contains six adjustable parameters, where
$\varepsilon=\varepsilon_{d}-\varepsilon_{_S}$ is the boson energy
and
$Q=(d^+\times\widetilde{s}+s^+\times\widetilde{d})^2+X(d^+\times\widetilde{d})^2$
here $X=CHI$. The parameters $a_0$, $a_1$, $a_2$, $a_3$ and $a_4$
designate the strength of the pairing, angular momentum,
quadrupole, octupole and hexadecapole interaction between the
bosons.

\section{Result and Discussion }

In this section, the calculated results can be discussed
separately for low -lying states of even-even isotopes of $Yb$,
with neutron number from 100 to 106. Our results include energy
levels and the reduced probability of $E2$-- transitions.

\subsection{Energy levels}

To describe the positive parity states in $PhM$, we determine the
model parameters via the following procedure. In accordance with
\cite{Bengtsson79}, we suppose that, at low spins, the rotational
core energy is the same as the energies of the ground band states.

Description of the parameters:

\noindent -- the inertial parameters $\jmath_{_0}$ and
$\jmath_{_1}$ of rotational core determined by (\ref{Eq.6}), using
the experimental energy of ground band states up to spin
$I\leq10\hbar$;

 \noindent -- headband energy of ground ($gr$)--,
$\beta_1$-- and $\beta_2$-- band states taken from experimental
data, because they are not perturbed by the Coriolis forces at
$I=0$;

\noindent -- headband energy of $\gamma$--, $K^{\pi}=1^+$ bands
($\omega_{\gamma}$ and $\omega_{1^+}$) and also matrix elements
$\langle K|j_x|K^{'}\pm1\rangle$ are free parameters of the model.
They have been fitted by the least squares method requiring the
best agreement between the theoretical energies and experimental
data. The fitting parameters of  model are given in Table 1.

\noindent Also, in the present work the rotational limit of the
$IBM-1$ has been applied to $^{170-176}Yb$, from the ratio
$\left(E\left(4^+\right)/E\left(2^+\right)\right)$ it has been
found that the $^{170-176}Yb$ isotopes are rotational (deformed
nuclei) and these nuclei have been successfully treated as
axhibiting the $SU(3)$ symmetry of $IBM-1$. The calculations have
been performed with the code $IBM-1.for$ and hence, no distinction
is made between neutron and proton bosons. For the analysis of
excitation energies in $Yb$ isotopes it we tried to keep to the
minimum the number of free parameters in Hamiltonian. The explicit
expression of Hamiltonian adopted in calculations is
\cite{Casten88}.
\begin{eqnarray}\label{Eq.11}
H=a_1L\cdot L+a_2Q\cdot Q
\end{eqnarray}

In framework of the $IBM-1$, the isotopic chains of $Yb$ with
$Z=70$ nuclei, having a number of proton bosons holes 6, a number
of neutron bosons particles varies from 9 to 11 for
$^{170-174}Yb$, and number of neutron boson hole for $^{176}Yb$ is
10. In Table 2 shown the coefficient values which we are using in
$IBM-1.for$. The comparison of calculated energy levels and
experimental data of low-lying states of $^{170}Yb$, $^{172}Yb$,
$^{174}Yb$ and $^{176}Yb$ isotopes are presented in the Fig. 1-4,
respectively. The $PhM$ calculations are plotted on the left of
the experimental data and $IBM-1$ calculations on the right of it
for $gr$--, $\beta_1$-- and $\gamma_1$-- band. The experimental
data are taken from \cite{Begzhanov89} for all isotopes of $Yb$
and also from \cite{Baglin02}-\cite{Basunia06} for $^{170-174}Yb$
and $^{176}Yb$, respectively. From these figures, we can see that
our calculated results (energies, spin and parity) in both models
are reasonably consistent with experimental data, except
$\gamma_1$-- band energies in $IBM-1$ calculations for $^{172}Yb$
and $^{174}Yb$ nuclei disagree with the experimental data. Also
the phenomenological calculations are in good agreement with the
experimental data than from those of $IBM-1$. In the high spin
these figures show that the difference between our calculation and
the experimental data. Furthermore the phenomenological model
predicts the energies, spin and parity of $\beta_2$--,
$\beta_3$--, $\beta_4$--, $\gamma_2$-- and $1^+$ band and as it
shown in the Tables 3-7, respectively. Finally, we believe that
the failure to use pairing parameter was the cause of the
disagreement between the $IBM-1$ calculations and experimental
data that will be discussed in future studies.

\subsection{The Reduced probability of $B(E2)$-- transitions}

In the $PhM$, with the wave functions calculated by solving the
Shr\"{o}dinger eq. (\ref{Eq.4}), the reduced probabilities of
$E2$-- transitions between states $I_iK_i$ and states $I_fK_f$ are
calculated \cite{Usmanov97,Usmanov10} as:
\begin{eqnarray}\label{Eq.12}
B(E2;I_iK_i\rightarrow
I_fK_f)&=&\left\{\sqrt{\frac{5}{16\pi}}eQ_0\left[\Psi_{gr,gr}^{I_f}
\Psi_{gr,K_i}^{I_i}C_{I_i0;20}^{I_f0}\right.\right. \nonumber \\
&+&\left.\left.\sum_{n}\Psi_{K_n,gr}^{I_f}
\Psi_{K_n,K_i}^{I_i}C_{I_iK_n;20}^{I_fK_n}\right]\right. \nonumber \\
&+&\left.
\sqrt{2}\left[\Psi_{gr,gr}^{I_f}\sum_{n}\frac{(-1)^{K_n}m_{_{K_n}}\Psi_{K_n,K_i}^{I_i}}{\sqrt{1+\delta_{K_n,0}}}
C_{I_iK_n;2-K_n}^{I_f0}\right.\right.
\nonumber \\
&+&\left.\left.
\Psi_{gr,K_i}^{I_i}\sum_{n}\frac{m_{_{K_n}}\Psi_{K_n,gr}^{I_f}}{\sqrt{1+\delta_{K_n,0}}}
C_{I_iK_n;2K_n}^{I_fK_n}\right]\right\}^2
\end{eqnarray}

\noindent where $m_{_{K_n}}=<gr|\hat{m}(E2)|K^{\pi}>$
$\left(K^{\pi}=0^+, 2^+\right.$ and $\left.1^+\right)$ are
constants to be determined from the experimental data, $Q_0$ is
the internal quadrupole moment of the nucleus, and
$C_{I_iK_i;2(K_i+K_f)}^{I_fK_f}$ and are the Clebsch-Gordon
coefficients.

Another advantage of the interacting $d$- boson model is that the
matrix elements of the electric quadrupole operator. The reduced
matrix elements of the $E2$ operator $\widehat{T}(E2)$ has the
form \cite{Arima76}-\cite{Arima79}
\begin{eqnarray}\label{Eq.13}
\widehat{T}(E2)&=&\alpha2\left[d^\dag \widetilde{s}+s^\dag
\widetilde{d}\right]^2+\beta2\left[d^\dag \widetilde{d}\right]^2
\\ \nonumber
&=&\alpha2\left\{\left[d^\dag \widetilde{s}+s^\dag
\widetilde{d}\right]^2+\chi\left[d^\dag
\widetilde{d}\right]^2\right\} =e_{_B}Q
\end{eqnarray}

\noindent here $\alpha2$ and $\beta2$ are two parameters and
$\beta2=\chi\alpha2, \alpha2=e_{_B}$ ($effective charge$). The
values of these parameters are presented in Table 3. Then the
$B(E2)$ values are given by
\begin{eqnarray}\label{Eq.14}
B\left(E2;J_i\rightarrow J_f\right)=\frac{1}{2J_i+1}|\langle
J_f\|\widehat{T}(E2)\|J_i\rangle|^2
\end{eqnarray}

For the calculations of the absolute $B(E2)$ values two parameters
$\alpha2$ and $\beta2$ of equation (\ref{Eq.13}) are adjusted
according to the experimental $B(E2;2_{gr}^+\rightarrow
0_{gr}^+)$. Table 8 shows the values of the parameters $\alpha2$
and $\beta2$, obtained in the present calculations. We present our
calculated results of the reduced probability of $E2$--
transitions of both models, and the comparison of calculated
values of $B(E2)$ transitions with experimental data \cite{http00}
are given in Table 9 for all nuclei of interest. In general, most
of the calculated results in both models are reasonably consistent
with the available experimental data, except for few cases that
deviate from the experimental data. As mentioned in Table 9 $PhM$
calculations are better than those of $IBM-1$ when compare with
the experimental data, except $B(E2;2_{gr}^+\rightarrow 0_{gr}^+)$
for $^{170}Yb$, $^{174}Yb$ and $^{176}Yb$,
$B(E2;6_{gr}^+\rightarrow 4_{gr}^+)$ for $^{172}Yb$,
$B(E2;4_{gr}^+\rightarrow 2_{gr}^+)$ and
$B(E2;14_{gr}^+\rightarrow 12_{gr}^+)$ for 174Yb and also
$B(E2;12_{gr}^+\rightarrow 10_{gr}^+)$ B(E2; ) for $^{170}Yb$.

\newpage
\section{Summary}

In this paper, energy levels and the reduced probability of $E2$--
transitions positive parity for $^{170-176}Yb$ isotopes with
neutron numbers between 100 and 106 have been calculated through
$PhM$ and $IBM-1$ calculations using the $IBM.for$ and $IBMT.for$
programs. The predicted low-lying levels $\left(gr-,
\beta_1-\right.$ and $\gamma_1-$ band$\left.\right)$ by $PhM$ are
in good agreement with the experimental data as compared with
those by $IBM-1$ for all nuclei of interest. In addition, the
$PhM$ is successful in predicting the $\beta_2-$, $\beta_3-$,
$\beta_4-$, $\gamma_2-$ and $1^+$-- band while $IBM-1$ fails.
Also, the $3^+$-- band is predicted by $IBM-1$ for $^{172}Yb$ and
$^{174}Yb$ nuclei. All calculations are compared with the
available experimental data. Also, the reduced probability of
$E2$-- transitions of $PhM$ calculations are better than those of
$IBM-1$ when compare with the experimental data, except
$B(E2;2_{gr}^+\rightarrow 0_{gr}^+)$ for $^{170}Yb$, $^{174}Yb$
and $^{176}Yb$, $B(E2;6_{gr}^+\rightarrow 4_{gr}^+)$ for
$^{172}Yb$, $B(E2;4_{gr}^+\rightarrow 2_{gr}^+)$ and
$B(E2;14_{gr}^+\rightarrow 12_{gr}^+)$ for 174Yb and also
$B(E2;12_{gr}^+\rightarrow 10_{gr}^+)$ B(E2; ) for $^{170}Yb$.

\section*{Acknowledgements}
This work has been financial supported by IIUM University Research
Grant (Type B) EDW B13-034-0919 and MOHE Fundamental Research
Grant Scheme FRGS13-077-0315. We thank the Islamic Development
Bank (IDB) for supporting this work under scholarships Nos.
36/11201905/35/IRQ/D31 and 37/IRQ/P30. The author A. A. Okhunov is
grateful to Prof. Ph.N. Usmanov for useful discussion and exchange
ideas.

%%%\section*{References}

%%%\end{enumerate}

\newpage
List of Table captions

\noindent {\bf Table 1} The used parameters of $PhM$ to calculate
energy of low excited states in $Yb$ isotopes.

\noindent {\bf Table 2} The used parameters of $IBM-1$ to
calculate energy of excited states in $Yb$ isotopes.

\noindent {\bf Table 3} The value of parameters obtained from
$\widehat{T}(E2)$ operator in program IBMT.for for calculate the
absolute values of $B(E2)$ in (in $eb$).

\noindent {\bf Table 4} The energy levels of $\beta_2$-- band of
$Yb$ isotopes (in $MeV$).

\noindent {\bf Table 5} The energy levels of $\beta_3$-- band of
$Yb$ isotopes (in $MeV$).

\noindent {\bf Table 6} The energy levels of $\beta_4$-- band of
$Yb$ isotopes (in $MeV$).

\noindent {\bf Table 7} The energy levels of $\gamma_2$-- band of
$Yb$ isotopes (in $MeV$).

\noindent {\bf Table 8} The energy levels of $1^+$-- band of $Yb$
isotopes (in $MeV$).

\noindent {\bf Table 9} The values of $B(E2)$-- transitions
isotopes of $Yb$ (in $W.u.$).

\newpage
\begin{table}
\caption{} \label{t1}
%\caption{The used parameters to
%calculate energy of excited states in $Yb$ isotopes.} \label{t1}
\begin{center}
\begin{tabular}{ccccccccc}
\hline
%\multicolumn{5}{c}{Option} & \multicolumn{6}{c}{Paper type} \\
%\hline

$A$ \ \ & \ \
   \ $(j_x)_{gr, 1}$ \  &  \  $(j_x)_{\beta_1, 1}$ \  &
   \ $(j_x)_{\beta_2, 1}$ \  &  \  $(j_x)_{\beta_3, 1}$ \  &  \  $(j_x)_{\beta_4, 1}$ \  &
   \ $(j_x)_{\gamma_1, 1}$ \   &    \  $(j_x)_{\gamma_2, 1}$ \   & $Q_0$  \\ \hline

 $170$ \ &  0.186 & 0.394 & 0.659 & 0.908 & -- &
0.728 & --  \ & \ 780 (30) \\
 $172$ \ &  0.275 & 0.978 & 0.718 & 0.110 & 0.300 & 0.325 &
0.210 \ & \ 791 (18)   \\
 $174$ \ &  0.185 & 0.400 & 0.250 & 0.150 & 0.200 & 0.085 &
0.100 \  & \  782 (24)  \\
 $176$ \ & 0.200 & 0.540 & 0.400 & 0.100 & - & 0.090 &
-- \  & \  759 (3)   \\ \hline
\end{tabular}

\noindent\footnotesize{Note: $(j_x)_{_{K', K}}$-- are matrix
elements of the Coriolis interactions and $Q_0$-- is intrinsic
quadrupole moment of the nucleus (in $fm^2$ units) taken from
\cite{Begzhanov89}.}
\end{center}
\end{table}
%%\vspace{0.5cm}

\begin{table}
\caption{} \label{t2}
%\caption{The Adopted values of the parameters used for IBM-1
%calculations. All parameters are given in $MeV$.} \label{t2}
\begin{center}
\begin{tabular}{cccc}
\hline
%\multicolumn{1}{c}{Option} & \multicolumn{1}{c}{Paper type} \\
%\hline
$A$ & $a_{_1}$ & $a_{_2}$ & $CHI$  \\ \hline
\ $170$ \ &  \ 0.0094 \ & \ -0.0120 \ &  \ -1.30  \ \\
\  $172$ \ &  \ 0.0091 \ & \ -0.0112 \ &  \ -1.20  \ \\
\  $174$ \ &  \ 0.0084 \ & \ -0.0150 \ &  \ -1.24  \ \\
\ $176$ \ &  \ 0.0089 \ & \ -0.0126 \ &  \ -1.30  \ \\
\hline
\end{tabular}
\end{center}
\end{table}

\begin{table}
\caption{} \label{t8}
%\caption{The parameters obtained from $T^{E2}$ operator in program
%$IBMT$ (in $eb$).} \label{t8}
\begin{center}
\begin{tabular}{cccc}
\hline \ $A$ \ & \ $\alpha2 $ \ & \ $\beta2$  \ \\ \hline
 $170$ \ & \ \ 0.1060 \ & \ \ -0.140  \\
 $172$ \ & \ \ 0.1037 \ & \ \ -0.137  \\
 $174$ \ & \ \ 0.0960 \ & \ \ -0.126  \\
 $176$ \ & \ \ 0.0980 \ & \ \ -0.129  \\
\hline
\end{tabular}
\end{center}
\end{table}

\begin{table}
\caption{} \label{t4}
%\caption{The energy levels of $\beta_2$-- band of $Yb$ isotopes
%(in $MeV$).} \label{t3}
\begin{center}
\begin{tabular}{ccccccccc}
\hline
  & \multicolumn{2}{c}{$^{170}Yb$} & \multicolumn{2}{c}{$^{172}Yb$}
 & \multicolumn{2}{c}{$^{174}Yb$} & \multicolumn{2}{c}{$^{176}Yb$}
  \\ \cline{2-3}\cline{4-5}\cline{6-7}\cline{8-9}
$I$ & Exp.\cite{Begzhanov89,Baglin02}  & $PhM$
 & Exp.\cite{Begzhanov89,Singh95}  & $PhM$
 & Exp.\cite{Begzhanov89,Browne99}  & $PhM$
 & Exp.\cite{Begzhanov89,Basunia06}  & $PhM$  \\ \hline
$0^+$ & 1.228   & 1.228 & 1.404   & 1.404
 & 1.885   & 1.885 & 1.518   & 1.518  \\

$2^+$ & 1.306  & 1.313  & 1.476   & 1.483
 & 1.958   & 1.962  & 1.610   & 1.601  \\

$4^+$ & --  & 1.507  & 1.632   & 1.666
 & 2.123   & 2.140  & --  & 1.792  \\

$6^+$ & --  & 1.804  & --  & 1.947
 & --  & 2.414  & --  & 2.086  \\

$8^+$ & --  & 2.195  & --  & 2.317
 & --  & 2.770  & --  & 2.476  \\

$10^+$ & --  & 2.669  & --  & 2.769
 & --  & 3.221  & --  & 2.954  \\

$12^+$ & --  & 3.220  & --  & 3.295
 & --  & 3.740  & --  & 3.512  \\
\hline
\end{tabular}
\end{center}
\end{table}

\begin{table}
\caption{} \label{t5}
%\caption{The energy levels of $\beta_3$-- band of $Yb$ isotopes
%(in $MeV$).} \label{t4}
\begin{center}
\begin{tabular}{ccccccccc}
\hline
  & \multicolumn{2}{c}{$^{170}Yb$} & \multicolumn{2}{c}{$^{172}Yb$}
 & \multicolumn{2}{c}{$^{174}Yb$} & \multicolumn{2}{c}{$^{176}Yb$}
  \\ \cline{2-3}\cline{4-5}\cline{6-7}\cline{8-9}
$I$ & Exp.\cite{Begzhanov89,Baglin02}  & $PhM$
 & Exp.\cite{Begzhanov89,Singh95}  & $PhM$
 & Exp.\cite{Begzhanov89,Browne99}  & $PhM$
 & Exp.\cite{Begzhanov89,Basunia06}  & $PhM$  \\ \hline

$0^+$ & 1.479   & 1.479 & 1.794   & 1.794
 & 2.113   & 2.110 & 1.779   & 1.779  \\

$2^+$ & 1.534  & 1.564  & 1.849   & 1.873
 & 2.172   & 2.178  & --   & 1.862  \\

$4^+$ & 1.667  & 1.758  & 1.975  & 2.056
 & 2.336   & 2.356  & --  & 2.053  \\

$6^+$ & --  & 2.055  & 2.156 & 2.156
 & --  & 2.630  & --  & 2.347  \\

$8^+$ & --  & 2.446  & --  & 2.707
 & --  & 2.993  & --  & 2.737  \\

$10^+$ & --  & 2.920  & --  & 3.159
 & --  & 3.437  & --  & 3.215  \\

$12^+$ & --  & 3.471  & --  & 3.685
 & --  & 3.956  & --  & 3.773  \\
\hline
\end{tabular}
\end{center}
\end{table}

\begin{table}
\caption{} \label{t6}
%\caption{The energy levels of $\beta_4$-- band of $Yb$ isotopes
%(in $MeV$).} \label{t5}
\begin{center}
\begin{tabular}{ccccc}
\hline
  & \multicolumn{2}{c}{$^{172}Yb$} & \multicolumn{2}{c}{$^{174}Yb$}
   \\ \cline{2-3}\cline{4-5}
$I$  & Exp.\cite{Begzhanov89,Singh95}  & PhM
 & Exp.\cite{Begzhanov89,Browne99}  & PhM  \\ \hline

$0^+$ & 1.894  & 1.894 & 2.821  & 2.821  \\

$2^+$ & 1.956  & 1.973 &  --    & 2.898  \\

$4^+$ & 2.100  & 2.156  & --    & 3.076  \\

$6^+$ & --     & 2.437  & --    & 3.350  \\

$8^+$ & --     & 2.807  & --    & 3.713  \\

$10^+$ & --    & 3.259  & --    & 4.157  \\

$12^+$ & --    & 3.785  & --    & 4.676  \\
\hline
\end{tabular}
\end{center}
\end{table}

\begin{table}
\caption{} \label{t7}
%\caption{The energy levels of $\gamma_2$-- band of $Yb$ isotopes
%(in $MeV$).} \label{t6}
\begin{center}
\begin{tabular}{ccccc}
\hline
  & \multicolumn{2}{c}{$^{172}Yb$} & \multicolumn{2}{c}{$^{174}Yb$}
   \\ \cline{2-3}\cline{4-5}
 $I$ & Exp.\cite{Begzhanov89,Singh95}  & $PhM$
 & Exp.\cite{Begzhanov89,Browne99}  & $PhM$  \\ \hline

$2^+$ & 1.608  & 1.619 & 2.728  & 2.727  \\

$3^+$ & 1.700  & 1.698 & 2.793  & 2.804  \\

$4^+$ & 1.803  & 1.802 & 2.882  & 2.905  \\

$5^+$ & 1.926  & 1.931 & --     & 3.031  \\

$6^+$ & 2.075  & 2.083 & --     & 3.179  \\

$7^+$ & --     & 2.257 & --     & 3.350  \\

$8^+$ & --     & 2.453 & --     & 3.542  \\

$9^+$ & --     & 2.669 & --     & 3.754  \\

$10^+$ & --    & 2.905 & --     & 3.986  \\

$11^+$ & --    & 3.159 & --     & 4.236  \\

$12^+$ & --    & 3.431 & --     & 4.505  \\

$13^+$ & --    & 3.720 & --     & 4.791  \\

$14^+$ & --    & 4.026 & --     & 5.093  \\
\hline
\end{tabular}
\end{center}
\end{table}

\begin{table}
\caption{} \label{t8}
%\caption{The energy levels of $1^+$-- band of $Yb$ isotopes (in
%$MeV$).} \label{t7}
\begin{center}
\begin{tabular}{ccccccccc}
\hline
  & \multicolumn{2}{c}{$^{170}Yb$} & \multicolumn{2}{c}{$^{172}Yb$}
 & \multicolumn{2}{c}{$^{174}Yb$} & \multicolumn{2}{c}{$^{176}Yb$}
  \\ \cline{2-3}\cline{4-5}\cline{6-7}\cline{8-9}
$I$ & Exp.\cite{Begzhanov89,Baglin02}  & $PhM$
 & Exp.\cite{Begzhanov89,Singh95}  & $PhM$
 & Exp.\cite{Begzhanov89,Browne99}  & $PhM$
 & Exp.\cite{Begzhanov89,Basunia06}  & $PhM$  \\ \hline

$1^+$ & 1.606 & 1.605 & 2.009 & 2.006
      & 1.624 & 1.605 & 1.819 & 1.818  \\

$2^+$ & 1.832 & 1.662 & 2.047 & 2.059
      & 1.674 & 1.657 & 1.867 & 1.874  \\

$3^+$ & --    & 1.746 & 2.109 & 2.138
      & 1.733 & 1.734 & --    & 1.956  \\

$4^+$ & --    & 1.856 & 2.193 & 2.242
      & 1.859 & 1.835 & --  & 2.065  \\

$5^+$ & --    & 1.993 & --  & 2.371
      & --    & 1.961 & --  & 2.200  \\

$6^+$ & --    & 2.153 & --  & 2.523
      & --    & 2.109 & --  & 2.359  \\

$7^+$ & --    & 2.337 & --  & 2.697
      & --    & 2.280 & --  & 2.542  \\

$8^+$ & --    & 2.544 & --  & 2.893
      & --    & 2.472 & --  & 2.749  \\

$9^+$ & --    & 2.771 & --  & 3.109
      & --    & 2.684 & --  & 2.977  \\

$10^+$ & --   & 3.018 & --  & 3.345
       & --   & 2.916 & --  & 3.227  \\

$11^+$ & --   & 3.285 & --  & 3.599
       & --   & 3.166 & --  & 3.496  \\

$12^+$ & --   & 3.569 & --  & 3.871
       & --   & 3.435 & --  & 3.785  \\

$13^+$ & --   & 3.871 & --  & 4.160
       & --   & 3.721 & --  & 4.093  \\

$14^+$ & --   & 4.190 & --  & 4.466
       & --   & 4.023 & --  & 4.419  \\
\hline
\end{tabular}
\end{center}
\end{table}

\begin{table}
\caption{} \label{t9}
%\caption{The values of $B(E2)$-- transitions isotopes of $Yb$ (in
%$W.u.$).} \label{t9}
\begin{center}
\begin{tabular}{ccccccc}
\hline
  & \multicolumn{3}{c}{$^{170}Yb$} & \multicolumn{3}{c}{$^{172}Yb$}
   \\ \cline{2-4}\cline{5-7}
 $I_iK_i\rightarrow I_fK_f$ & Exp.\cite{http00} & $PhM$
 & $IBM-1$
& Exp.\cite{http00} & \ $PhM$  \ & \ $IBM-1$  \\ \hline

$2_{gr}^+\rightarrow 0_{gr}^+$ & 201(6) & 216 &
 198.543 & 212(2) & 212 &  211.689   \\

$4_{gr}^+\rightarrow 2_{gr}^+$ & -- & 309 & 280.768 & 301(20) & 303 &  299.697   \\

$6_{gr}^+\rightarrow 4_{gr}^+$ & -- & 340 & 303.549 & 320(30) & 334 &  324.746   \\

$8_{gr}^+\rightarrow 6_{gr}^+$ & 360(30) & 356 & 309.178 & 400(40) & 350 &  331.835   \\

$10_{gr}^+\rightarrow 8_{gr}^+$ & 356(25) & 366 & 306.15 & 375(23) & 359 &  329.971   \\

$12_{gr}^+\rightarrow 10_{gr}^+$ & 268(21) & 372 & 296.956 & 430(60) & 366 &  322.160   \\

$14_{gr}^+\rightarrow 12_{gr}^+$ & -- & 377 & 283.181 & 394$_{-45}^{+60}$ & 370 &  309.724   \\

$16_{gr}^+\rightarrow 14_{gr}^+$ & -- & 381 & 265.349 & -- & 374 &  293.311   \\

$18_{gr}^+\rightarrow 16_{gr}^+$ & -- & 383 & 243.819 & -- & 376 &  273.310   \\

$20_{gr}^+\rightarrow 18_{gr}^+$ & -- & 386 & 218.751 & -- & 379 &
249.967   \\  \hline

& \multicolumn{3}{c}{$^{174}Yb$} & \multicolumn{3}{c}{$^{176}Yb$}
   \\ \cline{2-4}\cline{5-7}
 $I_iK_i\rightarrow I_fK_f$ & Exp.\cite{http00} & $PhM$
 & $IBM-1$
& Exp.\cite{http00} & \ $PhM$  \ & \ $IBM-1$  \\  \hline

$2_{gr}^+\rightarrow 0_{gr}^+$ & 201(7) & 205 & 199.908 & 183(7) & 184 &  182.916   \\

$4_{gr}^+\rightarrow 2_{gr}^+$ & 280(9) & 294 & 283.321 & 270(25)& 263 &  258.969   \\

$6_{gr}^+\rightarrow 4_{gr}^+$ & 370(50) & 324 & 307.532 & 298(22) & 290 &  280.618   \\

$8_{gr}^+\rightarrow 6_{gr}^+$ & 388(21) & 339 & 315.122 & 300(5) & 303 &  286.743   \\

$10_{gr}^+\rightarrow 8_{gr}^+$ & 335(22) & 348 & 314.533 & -- & 312 &  285.139   \\

$12_{gr}^+\rightarrow 10_{gr}^+$ & 369(23) & 354 & 308.624 & -- & 317 &  278.384   \\

$14_{gr}^+\rightarrow 12_{gr}^+$ & 320(8) & 359 & 298.941 & -- & 321 &  267.636   \\

$16_{gr}^+\rightarrow 14_{gr}^+$ & -- & 362 & 285.192 & -- & 324 &  253.459   \\

$18_{gr}^+\rightarrow 16_{gr}^+$ & -- & 365 & 268.659 & -- & 327 &  236.177   \\

$20_{gr}^+\rightarrow 18_{gr}^+$ & -- & 367 & 249.249 & -- & 328 &
215.995
\\ \hline
\end{tabular}
\end{center}
\end{table}

\newpage
List of figure captions

{\bf Figure 1} The comparison of calculation energy values by
$PhM$ and $IBM-1$ with experimental data for $^{170}Yb$.

{\bf Figure 2} The comparison of calculation energy values by
$PhM$ and $IBM-1$ with experimental data for $^{172}Yb$.

{\bf Figure 3} The comparison of calculation energy values by
$PhM$ and $IBM-1$ with experimental data for $^{174}Yb$.

{\bf Figure 4} The comparison of calculation energy values by
$PhM$ and $IBM-1$ with experimental data for $^{176}Yb$.

\newpage
\begin{figure}
\begin{center}
\includegraphics[width=20pc]{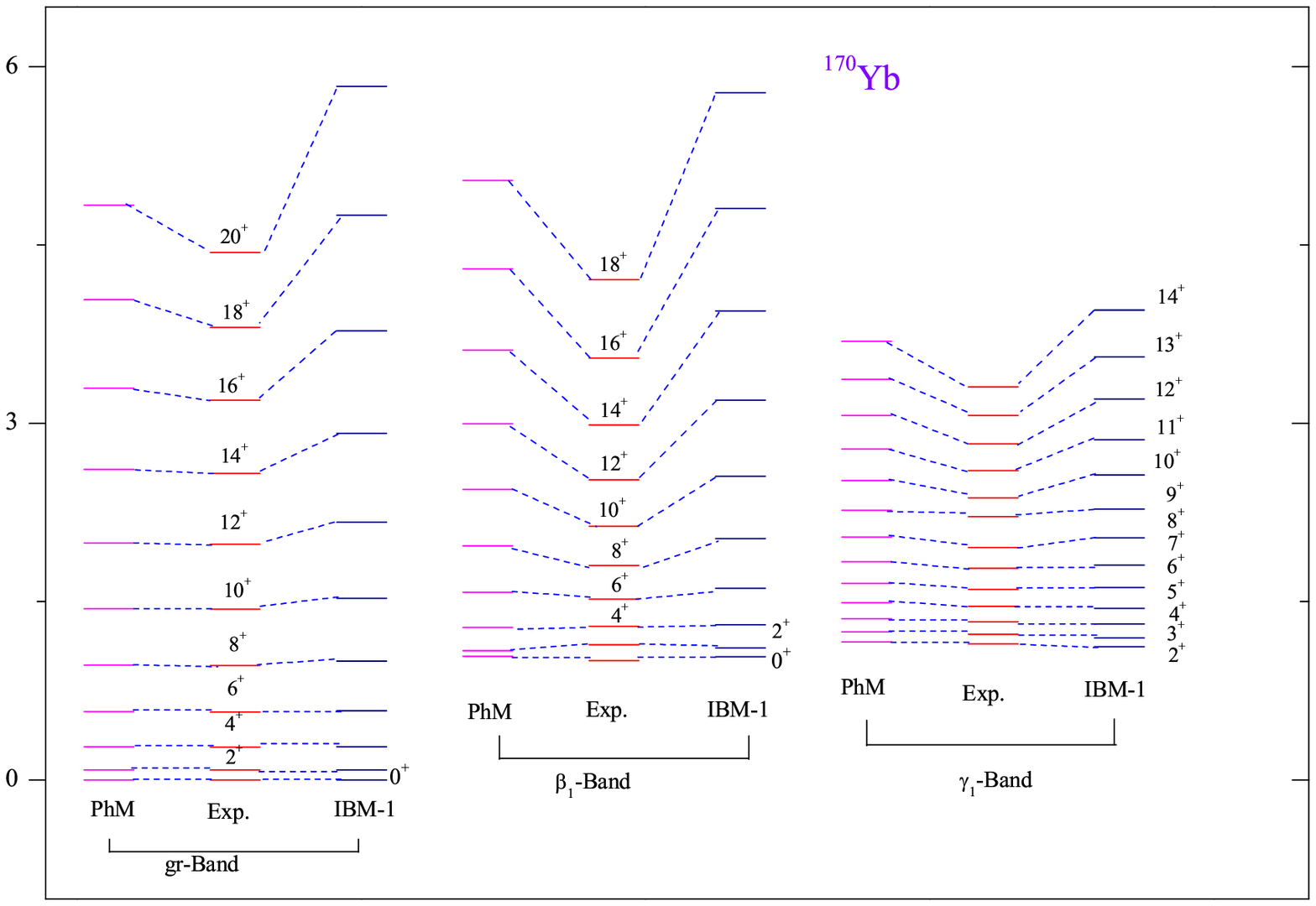}
\end{center}
\caption{} \label{f1}
%\caption{The comparison of calculation energy values by $PhM$ and
%$IBM-1$ with experimental data for $^{170}Yb$.} \label{f1}
\end{figure}

\begin{figure}
\begin{center}
\includegraphics[width=20pc]{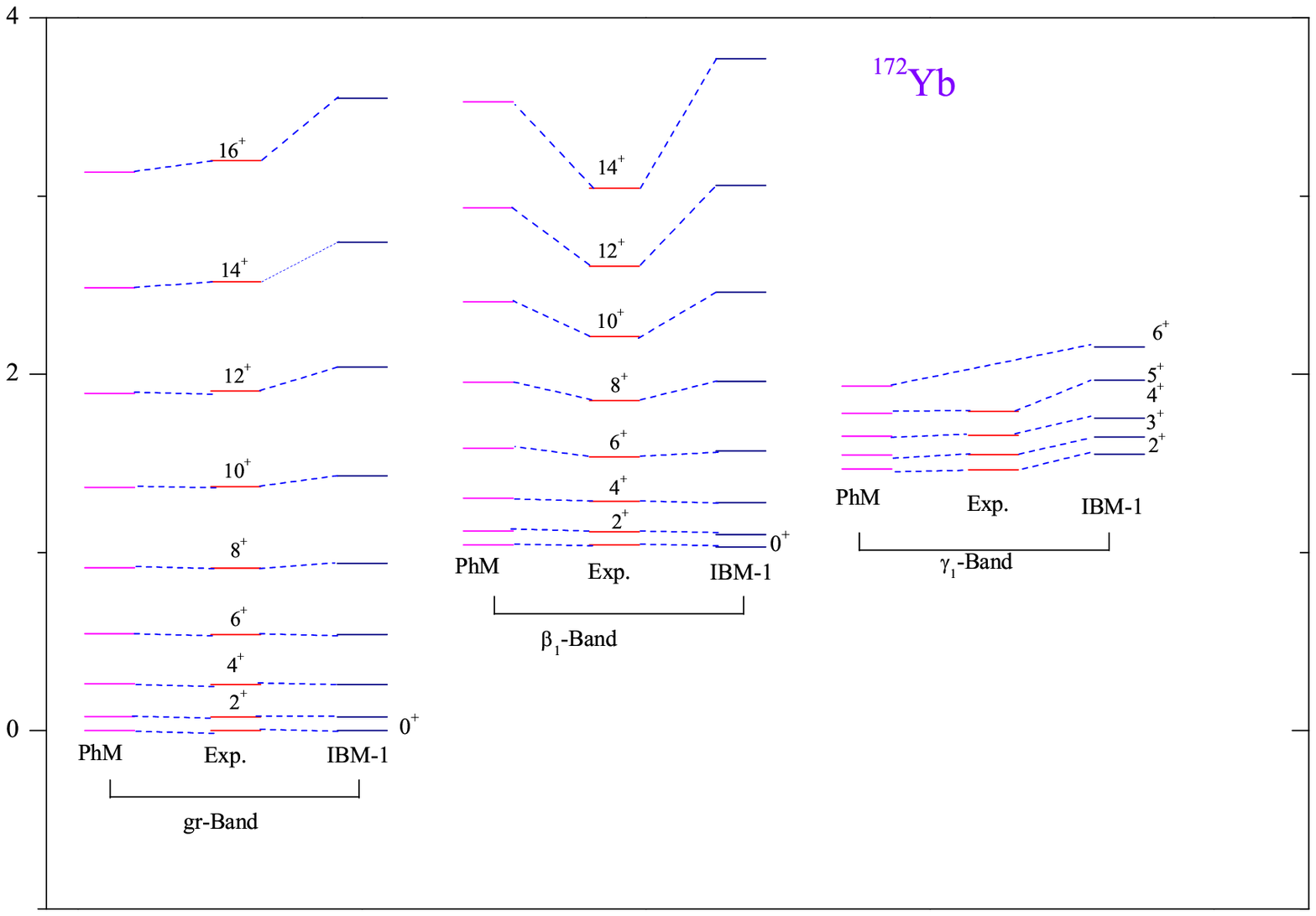}
\end{center}
\caption{} \label{f2}
%\caption{The comparison of calculation energy values by $PhM$ and
%$IBM-1$ with experimental data for $^{172}Yb$.} \label{f2}
\end{figure}

\begin{figure}
\begin{center}
\includegraphics[width=20pc]{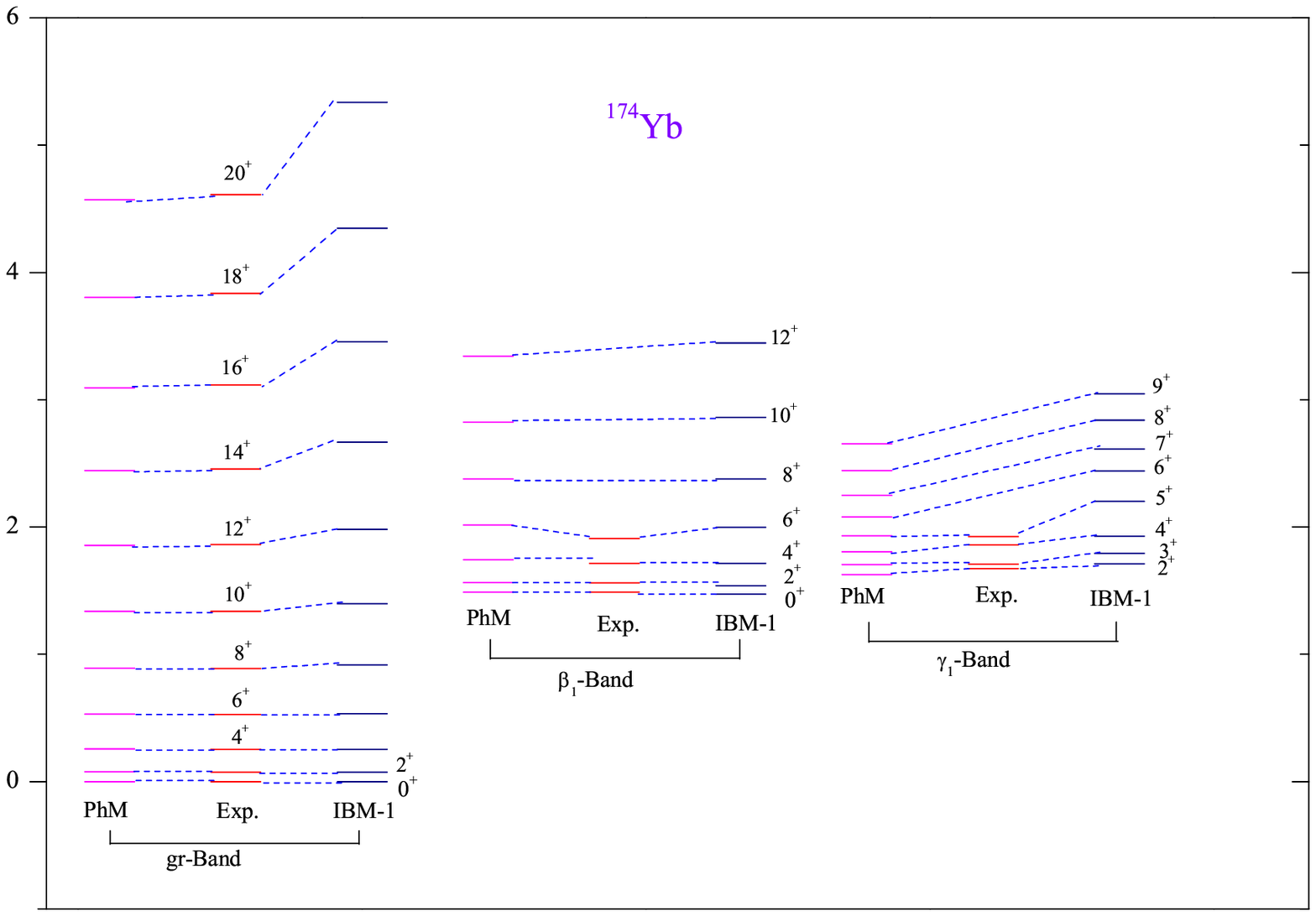}
\end{center}
\caption{} \label{f3}
%\caption{The comparison of calculation energy values by $PhM$ and
%$IBM-1$ with experimental data for $^{174}Yb$.} \label{f3}

\end{figure}

\begin{figure}
\begin{center}
\includegraphics[width=20pc]{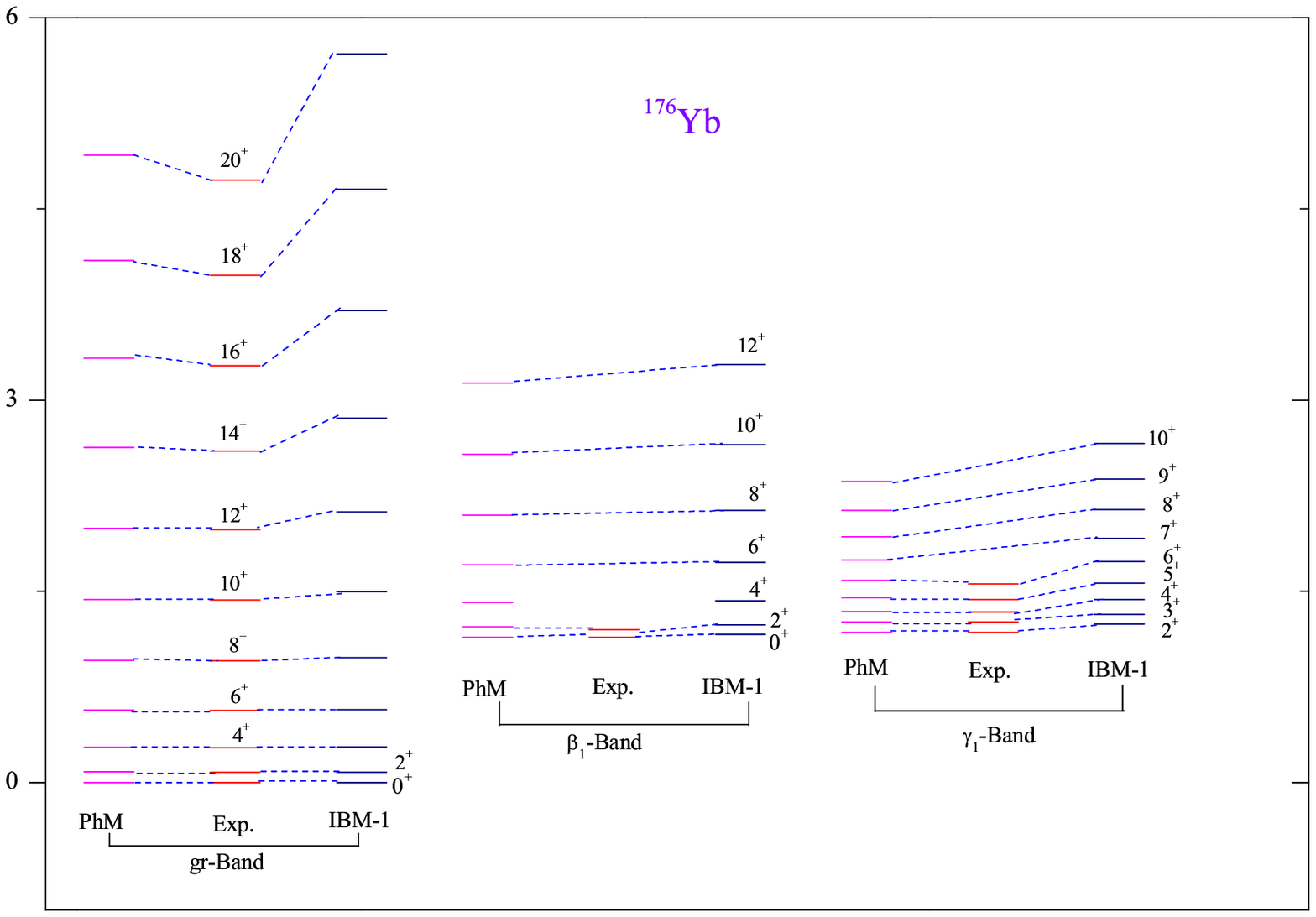}
\end{center}
\caption{} \label{f4}
%\caption{The comparison of calculation energy values by $PhM$ and
%$IBM-1$ with experimental data for $^{176}Yb$.} \label{f4}
\end{figure}

\end{document}